# Enhanced Electron Photoemission by Collective Lattice Resonances in Plasmonic Nanoparticle-Array Photodetectors and Solar Cells

Sergei V. Zhukovsky • Viktoriia E. Babicheva • Alexander V. Uskov • Igor E. Protsenko • Andrei V. Lavrinenko

**Abstract** We propose to use collective lattice resonances in plasmonic nanoparticle arrays to enhance and tailor photoelectron emission in Schottky-barrier photodetectors and solar cells. We show that the interaction between narrow-band lattice resonances (the Rayleigh anomaly) and broader-band individual particle excitations (localized surface plasmon resonances) leads to stronger local field enhancement. In turn, this causes significant increase of the photocurrent compared to the case when only individual particle excitations are present. The results can be used to design new photodetectors with highly selective, tunable spectral response, which are able to detect photons with the energy below the semiconductor bandgap. The findings can also be used to develop solar cells with increased efficiency.

## Introduction

Plasmonic nanostructures hold great promise in the design of advanced photodetectors and photovoltaic devices [1-3]. In particular, excitation of plasmonic resonances in metallic nanoantennas and electron photoemission from them was shown to extend the spectral response of photodetectors to the energies below the semiconductor bandgap edge [2,4-6]. Photoelectron emission from plasmonic nanoparticles can generate a photocurrent in addition to that originating from direct band-to-band absorption in semiconductor, increasing the device efficiency and putting forth a new concept of photoconductive metamaterials [2,5,8-10]. Electron photoemission can be enhanced by optimizing the individual nanoparticle shape and/or the Schottky barrier configuration at metal-semiconductor interfaces [2,4,8].

On the other hand, when metal particles are properly arranged in a lattice, *collective* effects become important, offering a further enhancement of local fields in the resonant plasmonic modes. Collective effects result from coherent interference of individual nanoparticle excitations in the far-field zone due to the dipole-dipole interaction between the nanoparticles [11-16]. These effects are strongly resonant, since they occur when the wavelength of incident light is close to the lattice period, satisfying the condition for the Rayleigh anomalies (RAs) in diffractive gratings [9,10,17,18]. The resonance conditions also give new possibilities of self-organizing process of Au film fragmentation and nanoparticles formation [19].

It was shown earlier that strong dipole interactions in a periodic array of large nanoparticles produce strong optical fields near their surface, up to orders of magnitude higher than for isolated nanoparticles of the same size and shape [20]. Furthermore, when the localized surface plasmon resonance (LSPR) frequency of an individual particle is close to the frequencies of RAs, the collective effects lead to narrow-band resonances with a Fano-like profile and quality factor much larger compared to that of the single-particle LSPR [18,21]. This effect can be utilized to enhance performance of various devices, e.g., sensors [22] and light sources [23]. Similar resonant effects were also very recently observed in subwavelength one-dimensional diffraction gratings, facilitating narrow-band photodetection [6].

In this paper, we report how collective effects can be exploited to increase electron photoemission from the nanoparticles in an array as compared to isolated nanoparticles. In particular, we show that interaction of narrow-band RAs with broader-band LSPR of individual nanoparticles results in narrow, Fano-shaped, highly tunable photoemission resonances in the array. These effects can be useful in the design of new narrow-band frequency- and polarization-selective photodetectors in the infrared range and in further improvement of plasmon-assisted photovoltaic devices by extending their operating spectrum beyond the band-to-band transition spectrum in semiconductors.

S. V. Zhukovsky (✉) • V. E. Babicheva • A. V. Uskov • A. V. Lavrinenko
DTU Fotonik – Department of Photonics Engineering, Technical University of Denmark,
Ørsteds Pl. 343, DK-2800 Kgs. Lyngby, Denmark
e-mail: sezh@fotonik.dtu.dk

A. V. Uskov • I. E. Protsenko
P. N. Lebedev Physical Institute, Russian Academy of Sciences,
Leninskiy Pr. 53, 119333 Moscow, Russia
Advanced Energy Technologies Ltd,
Skolkovo, Novaya Ul. 100, 143025 Moscow Region, Russia

V. E. Babicheva
Birck Nanotechnology Center, Purdue University,
1205 West State Street, West Lafayette, IN, 47907-2057 USA

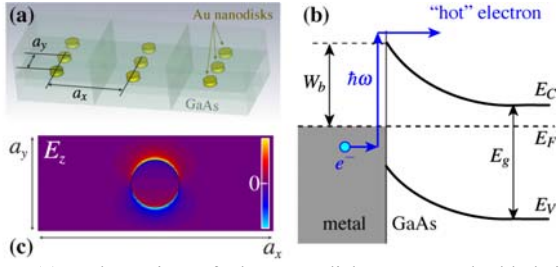

**Fig. 1** (a) Schematics of the nanodisk array embedded in uniform GaAs. (b) Schematics of the Schottky barrier at the Au/GaAs interface. (c) Local field enhancement (z-component of the electric field plotted for the at the bottom facet of the nanodisk) at the absorption peak (see Fig. 2b) for $a_x = 400$ nm.

## Resonant Absorption in Plasmonic Nanoparticle Arrays

We consider an array of metallic (Au) nanodisks with radius $r$ and thickness $h$, embedded into a semiconductor (GaAs) matrix with refractive index $n_m$ (Fig. 1a) on which light of frequency $\omega$ is incident normally. We are interested in the regime when

$$W_b < \hbar\omega < E_g, \qquad (1)$$

where $E_g$ is the band gap energy for the semiconductor matrix, and $W_b$ is the work function for the metal/semiconductor interface. If $\hbar\omega < E_g$, no photocurrent is generated in semiconductor due to photon absorption in band-to-band transitions. Nevertheless, if $\hbar\omega > W_b$, photocurrent can result from photoemission of "hot" electrons from metal nanoparticles into the semiconductor matrix across the Schottky barrier (Fig. 1b). For gold nanodisks in a GaAs matrix ($E_g = 1.43$ eV) the work function is $W_b \sim 0.8$ eV (with image force correction). Thus, Eq. (1) is fulfilled for $\hbar\omega$ from 0.8 to 1.43 eV, which corresponds to wavelengths between 870 and 1550 nm.

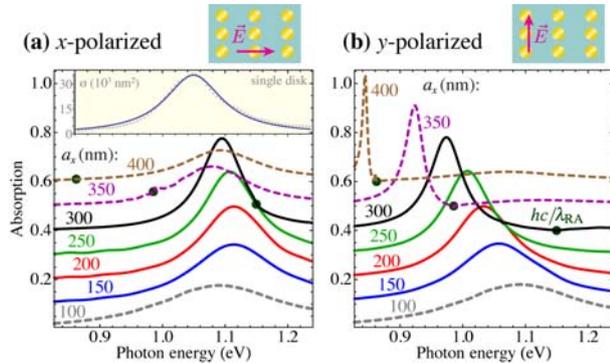

**Fig. 2** Calculated absorption spectra of nanodisk lattices with $a_y = 100$ nm and varying $a_x$ for light polarized along (a) x-axis and (b) y-axis. The dots show the energy corresponding to $\lambda_{RA}$ for each $a_x$. The plots are separated by 0.1 in the y-axis to improve readability. The inset in (a) shows the absorption cross-section of a single nanodisk (a Lorentzian fit to time-domain calculations).

Each single nanodisk can exhibit an LSPR at frequency $\omega_{LSPR}$. If $\hbar\omega$ fulfils Eq. (1) and the frequency $\omega$ is close to $\omega_{LSPR}$, "hot" electron photoemission becomes resonantly plasmon-enhanced [2,5,8].

Furthermore, the nanoparticle lattice (Fig. 1a) can be viewed as a two-dimensional (2D) diffraction grating with periods $a_x$ and $a_y$, where RAs occur at wavelengths $\lambda_{RA}^{(m_x,m_y)}$, for which evanescent diffracted waves become propagating. In metallic gratings, RAs can be regarded as wavelengths where the grating couples the incident light to a surface plasmon-polariton wave propagating in the grating plane [15,17-19]. For normal incidence of light, $\lambda_{RA}^{(m_x,m_y)}$ are determined by [17]

$$\frac{2\pi n_m}{\lambda_{RA}^{(m_x,m_y)}} = \sqrt{\left(\frac{2\pi m_x}{a_x}\right)^2 + \left(\frac{2\pi m_y}{a_y}\right)^2}. \qquad (2)$$

The collective lattice resonances become significantly pronounced once $\lambda_{RA}^{(m_x,m_y)}$ approaches $\lambda_{LSPR} = 2\pi/\omega_{LSPR}$. Thus we aim at designing a 2D grating with $\lambda_{RA}^{(m_x,m_y)}$ close to $\lambda_{LSPR}$. It is important to stress that a refractive index step close to the lattice plane suppresses collective effects [12,13,24]. For this reason, we consider complete embedding of the nanoparticles into the semiconductor matrix. In addition, embedding expands the area of the metal-semiconductor contact, thereby increasing the photocurrent.

In dense lattices [2,4,8], where both $a_x$ and $a_y$ are less than $\lambda_{LSPR}/n_m$, the RA wavelengths (2) are far from the operating wavelength range $\lambda \sim \lambda_{LSPR}$ and the LSPR peak is the dominant feature in the absorption spectra. In contrast to that, we consider a rectangular lattice and only keep it dense in one (y) direction, making it sparser in the other (x) direction (see Fig. 1a). Thus, we fix $a_y$ at a constant small value and vary $a_x$ up to the values for which the greatest RA wavelength

$$\lambda_{RA} = \lambda_{RA}^{(1,0)} = a_x n_m \qquad (3)$$

approaches $\lambda_{LSPR}$. Varying $a_x$ in order to move $\lambda_{RA}$ towards $\lambda_{LSPR}$ complements another mechanism of bringing these two resonances together, namely, moving $\lambda_{LSPR}$ towards $\lambda_{RA}$ by changing the nanoantenna size and geometry, also shown to result in narrow-band absorption [18].

Note that RAs are lattice effects related to coherent coupling between many nanoparticles distributed over a distance. Such coupling occurs through the long-range terms in the dipole radiation of each individual scatterer [11], which vanishes in the direction parallel to the induced dipole moment (i.e., polarization of incident light). Therefore, plasmonic properties of a rectangular nanoparticle lattice should be most sensitive to variation of the lattice constant in the direction perpendicular to the light polarization under normal incidence. Specifically, if we vary $a_x$, we expect stronger and narrower plasmonic resonances (and hence, greater capabilities for photoemission enhancement) for the *y*-polarization of incident light.

As an example we considered the structure similar to the one studied earlier [8], with $r = 25$ nm, $h = 18$ nm, $n_m = 3.6$, $a_y = 100$ nm, and $a_x$ varied between 100 and 400 nm. Permittivity of gold was described by the Drude model with plasma frequency $2.18\times10^{15}$ s$^{-1}$ and collision frequency $6.47\times10^{12}$ s$^{-1}$ [25]. Full-wave numerical calculations were carried out in the frequency domain using CST Microwave Studio [26].

The calculated absorption spectra are shown in Fig. 2. For the dense lattice ($a_x = a_y = 100$ nm), we note that the absorption spectrum is identical for both polarizations and closely resembles the absorption cross-section of an individual nanoparticle (Fig. 2a, inset), in full accordance with prior knowledge [8]. As expected, the LSPR resonance is broadened due to there being many particles in a lattice. For light polarized along the *x*-axis (Fig. 2a), there is a slight narrowing of the LSPR absorption peak and a very slight shift of its wavelength with changing $a_x$ so that the maximum of response is close to the resonance wavelength $\lambda_{LSPR}$ of an individual particle (as it occurs in the dense array [8]) even if $\lambda_{RA}$ approaches $\lambda_{LSPR}$.

However, for light polarized along the *y*-axis (Fig. 2b), the effect of the lattice resonances becomes much stronger in the spectral response. We see that absorption turns to zero at $\hbar\omega = 2\pi\hbar c/\lambda_{RA}$ (Fig. 2b, dots), where higher-order diffraction appears [15]. We also see a sharp peak in the absorption spectra, apparently associated with the interaction of a narrow-band lattice RA resonance with a broader LSPR of the individual nanoparticle. This absorption peak follows $\lambda_{RA}$ towards lower frequencies and acquires an asymmetric Fano-like shape for larger $a_x$. Finally, it can be noticed that the individual-particle LSPR response reappears in the absorption spectra when $\lambda_{RA}$ becomes significantly larger than $\lambda_{LSPR}$. For even larger $a_x$ the individual-particle response is expected to resume [16].

**Photoemission on Metal-Semiconductor Interface**

Calculating the field distribution at the absorption peak frequency reveals that the strongest local fields are located in the vicinity of the nanodisk surface (see Fig. 1c). This field enhancement underlies the enhanced electron photoemission from the nanoparticles.

Two different physical mechanisms of electron photoemission can be defined [27-29]. The first one is absorption of a photon by an electron as it collides with the nanoparticle boundary, causing electron emission from the metal (the surface photoelectric effect). The second mechanism is absorption of a photon by an electron *inside* the nanoparticle (the volume photoelectric effect) with subsequent transport of the "hot" electron to the surface and its emission over the Schottky barrier [30]. While the question on which of these two mechanisms is prevalent in each particular metallic structure is still open, the crucial role of strong field enhancement near the metal surface is undoubted.

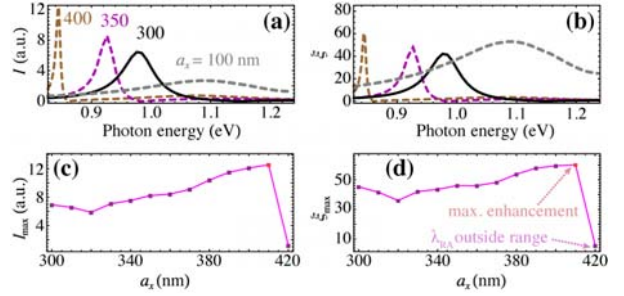

**Fig. 3** Spectral dependence of (a) one-particle photocurrent $I_{NP}$ and (b) photocurrent enhancement factor $\xi$ for different values of the lattice period $a_x$. Also shown is the dependence of (c) $I_{max}(a_x)$ and (d) $\xi_{max}(a_x)$. Incident light is *y*-polarized.

Assuming that the surface-driven effect prevails over the bulk effect in small nanoparticles [27], a detailed theory of photoemission from plasmonic nanoparticles was developed [5,8]. If particle sizes are larger than the de Broglie wavelength of the electron, the photocurrent from such nanoparticle is proportional to the squared normal component of the electric field $E_n$ integrated over the particle (disk) surface [5,8]:

$$I_{NP}(\lambda) = C_{em}(\lambda)\oint_{disk}|E_n|^2 dS . \qquad (4)$$

The proportionality coefficient $C_{em}(\lambda)$ depends on the properties of the Schottky barrier between metal and semiconductor, and in particular, on the work function $W_b$ [5,30]. The photocurrent density per unit area of a photodetector device is then

$$J_{device} = \frac{I_{NP}}{a_x a_y} = C_{em}(\lambda)|E_0|^2 \xi , \qquad (5)$$

where the dimensionless quantity

$$\xi = |E_0|^{-2} a_x^{-1} a_y^{-1} \oint_{disk}|E_n|^2 dS \qquad (6)$$

is the field enhancement factor relatively to the incident field $E_0$. The intensity of the incident light in GaAs matrix is $S = cn_m|E_0|^2/8\pi$ so that the quantum efficiency $\eta = (J_{device}/e)/(S/\hbar\omega)$ of photoemission from the nanoparticle array can be expressed as

$$\eta = \eta_0 \cdot \xi , \quad \eta_0 = \frac{8\pi\hbar\omega C_{em}}{n_m ec} . \qquad (7)$$

The dimensionless coefficient $\eta_0$ is about $10^{-5}$ to $10^{-3}$ depending on material parameters and wavelength for the chosen mechanism of photoemission [5,8]. It can be considered as the quantum yield of conventional

Au/GaAs Schottky barriers photodetector with a flat interface. Small value of $\eta_0$ usually limits practically useful applications of such conventional photodetectors to the determination of barrier height $W_b$. However, if the plasmonic enhancement factor $\xi$ is large enough, the overall quantum yield is expected to reach values that would permit a wider range of applications.

**Photoemission Enhancement Due to Collective Effects**

Calculating the photocurrent using Eqs. (4)–(7) with the numerically obtained field profiles shows that increased absorption is indeed accompanied by corresponding increase in the photocurrent. Thus, an enhanced absorption resonance resulting from coupling effects in the nanoparticle lattice (Fig. 2b) gives rise to a sharp spectral peak both in the photocurrent from an individual nanoparticle $I_{NP}$ (Fig. 3a) and in the device photocurrent enhancement factor $\xi$ (Fig.3b).

Specifically, comparing the shape of $I_{NP}(\hbar\omega)$ for different $a_x$ reveals that the photoemission peak becomes much higher but narrower for larger $a_x$. As an example, increasing $a_x$ from 300 to 400 nm nearly doubles the peak value of $I_{NP}$ (Fig. 3c). Compared to the reference dense lattice with $a_x = a_y = 100$ nm, where, in fact, only the individual-particle resonances exist in the operating frequency range (1), the total photocurrent from one particle increases almost by a factor of three.

Although the maximum of current $I_{NP}$ from a nanoparticle rises very substantially when $a_x$ is increased from 100 to 400 nm, the rise in the spectral maximum of the "macroscopic" property $J_{device} = I_{NP}/(a_x a_y) \propto \xi$, as well as in $\eta \propto \xi$, is more modest [see Fig. 3d for $\xi(a_x)$]. It is because the increase in the photoemission from *one* nanoparticle due the lattice resonances is partially compensated by the decrease in the nanoparticle density: $\xi \propto 1/(a_x a_y)$. Thus, one can achieve the same photocurrent densities with a much sparser lattice of nanoparticles than previously reported [8]. Additionally, Figs. 3b,d indicate that the quantum yield of the nanoparticle array $\eta$ can be between 0.05 and 10%, depending on the value of $\eta_0$. The accurate estimation of the overall quantum yield, as well as its enhancement by optimizing the geometrical parameters of the nanoparticles, is a subject of forthcoming research.

As mentioned above and in earlier works [7,8], the resonant photoemission for photons with energies below the semiconductor bandgap edge can be used to increase the solar cell efficiency. This concept would work only if reflection and absorption in the nanoparticle array do not reduce effective absorption of solar visible-range photons in semiconductor. In this respect, a sparse nanoparticle lattice (with the coverage of ~5%) is preferable to denser arrays [2,8], let alone grating-like structures [6] where the coverage of the device surface is between 69 and 77%. Estimations show that increasing $a_x$ from 100 to 400 nm increases the average visible-range ($1.43 < \hbar\omega < 3$ eV) transmittance of the particle array from 89% to 97% and decreases the undesired light absorption by the particles from 9% to 3%.

One has to keep in mind that as photoelectrons are emitted by an electrically insulated nanoparticle, the charge build-up will raise the potential barrier at the particle surface, which may eventually prevent further electrons from being emitted. To counteract this effect, there should be a mechanism to replenish the emitted electron in the nanoparticle. For discrete particles (i.e. without direct electric contact), one possible approach is to cover the nanoparticle array with a layer of transparent conductive oxide (TCO), which provides an electric contact with an external circuit, from which electrons can flow to the nanoparticles [2,8].

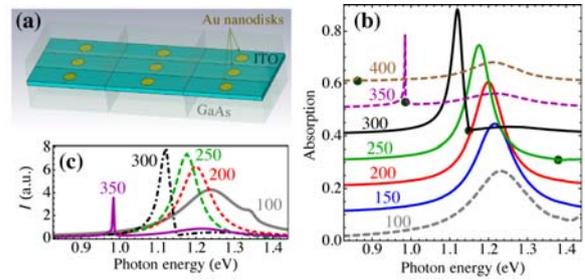

**Fig. 4** (a) Schematics, (b) absorption spectra (same as Fig. 2b, separated by 0.1 in the *y*-axis to improve readability), and (c) photocurrent spectra (same as Fig. 3a) for the modified design where the nanodisk array is embedded in a transparent conductive oxide film (ITO) and sandwiched between GaAs layers, providing an Au/GaAs interface at the front and back facets of the nanodisk. Incident light is *y*-polarized.

However, as was mentioned above, the mismatch between refractive indices of TCO and GaAs is detrimental to collective resonances in the nanoparticle array [13]. To remedy this, we consider a symmetric structure, where nanoparticles are embedded in a thin TCO layer and sandwiched between semiconductor layers (Fig. 4a). The electric contact to replenish emitted electrons in nanoparticles is provided through the side walls, whereas photoemission itself happens at the Schottky barriers at the front and back facets of the nanodisks.

Calculation results for the TCO material taken to be indium tin oxide (ITO) annealed in $N_2$ at 450°C [31] are shown in Fig.4b-c. It can be seen that the absorption and photoemission spectra show similar features as for completely embedded nanoparticles (the design from Fig. 1a). The peak values of both absorption and photocurrent are seen to be slightly lower, especially for $a_x > 300$ nm, when the resonance becomes extremely narrow-band. Still, the contribution of lattice resonances to photoemission enhancement in the sparse array configuration remains strong, so the alternative design of Fig. 4 can provide increased photoemission as well.

**Conclusions and Outlook**

In conclusion, we report that collective effects in plasmonic nanoparticle arrays can result in enhanced photoemission of electrons from nanoparticles embedded into Schottky-barrier photodetectors. The lattice resonances lead to a strong narrowing of the absorption spectral peak and increasing of the electron photoemission from each nanoparticle by several times. As a result one can achieve the same peak photocurrent densities (or, similarly, the same high levels of the quantum efficiency of photoemission) as in the previously reported dense arrays [8] but with a much lower density of nanoparticles. Owing to the dipole-dipole coupling between the particles in the lattice, a greater degree of control over photoelectron emission can be achieved, resulting in narrow-band photoemission with strong potential for tunability. In this respect nanoparticles can be regarded as nanoantennas, and the proposed *coupled nanoantenna arrays* can be useful in the design of photodetectors with high sensitivity, selectivity, and efficiency. The photoemission spectra can be tuned and optimized by changing the period of the structure, as well as the size and shape of the nanoparticles. One promising direction is moving from single nanodisk lattices to lattices of nanodisk clusters, as was done for nanodisk dimers in Ref. [18].

The proposed nanoparticle arrays can also be useful in photovoltaic applications. It was already suggested [7,8] that hot-electron photoemission from plasmonic nanoantennas can increase the efficiency of solar cells by harvesting the energy of photons in the solar spectrum, which are unable to produce photocurrent via band-to-band transitions in semiconductors but can cause electron photoemission from metal over a Schottky barrier. In this respect, the implications of our results are twofold. On the one hand, denser lattices with a broader absorption peak are more preferable from the point of view of the efficiency across the infrared range, because a broad response allows more photons to be harvested. On the other hand, sparser arrays are more favorable for inclusion into hybrid photovoltaic elements combining band-to-band and plasmon-assisted photocurrent generation. There, lower particle density reduces reflection losses in the visible range, which may benefit the overall efficiency. Again, a controlled coupling between the individual particles (antennas) in the array is a useful tool to tailor photovoltaic elements to the particular needs.

**Acknowledgments** S.V.Z. acknowledges financial support from the People Programme (Marie Curie Actions) of the European Union's 7th Framework Programme FP7-PEOPLE-2011-IIF under REA grant agreement No. 302009 (Project HyPHONE). "I.E.P. and A.V.U. acknowledge support from the Russian Foundation for Basic Research (Project No. 14-02-00125) and the Russian MSE State Contract N14.527.11.0002, and from the CASE project (Denmark).

**References**

1. Atwater H, Polman A (2010) Plasmonics for improved photovoltaic devices. Nat Mater 9:205–213
2. Knight MW, Sobhani A, Nordlander P, Halas NJ (2011) Photodetection with active optical antennas. Science 332:702–704
3. Akimov YA, Koh WS (2011) Design of plasmonic nanoparticles for efficient subwavelength light trapping in thin-film solar cells. Plasmonics 6(1):155–161
4. Knight MW, Wang Y, Urban AS, Sobhani A, Zheng BY, Nordlander P, Halas NJ (2013) Embedding plasmonic nanostructure diodes enhances hot electron emission. Nano Lett 13:1687–1692
5. Protsenko IE, Uskov AV (2012) Photoemission from metal nanoparticles. Phys Usp 55(5):508–518
6. Sobhani A, Knight MW, Wang Y, Zheng B, King NS, Brown LV, Fang Z, Nordlander P, Halas NJ (2013) Narrowband photodetection in the near-infrared with a plasmon-induced hot electron device. Nature Comm 4(3):1643
7. Moulin EA, Paetzold UW, Pieters BE, Reetz W, Carius R (2013) Plasmon-induced photoexcitation of "hot" electrons and "hot" holes in amorphous silicon photosensitive devices containing silver nanoparticles. J Appl Phys 113(14):144501
8. Novitsky A, Uskov AV, Gritti C, Protsenko IE, Kardynał BE, Lavrinenko AV (2012) Photon absorption and photocurrent in solar cells below semiconductor bandgap due to electron photoemission from plasmonic nanoantennas. Prog Photovolt Res Appl. doi: 10.1002/pip.2278
9. White TP and Catchpole KR (2012) Plasmon-enhanced internal photoemission for photovoltaics: Theoretical efficiency limits. Appl. Phys. Lett. 101, 073905
10. Govorov AO, Zhang H, and Gun'Ko YK (2013) Theory of Photoinjection of Hot Plasmonic Carriers from Metal Nanostructures into Semiconductors and Surface Molecules. J. Phys. Chem. C 117 (32):16616–16631
11. Zou S, Schatz GC (2006) Theoretical studies of plasmon resonances in one-dimensional nanoparticle chains: narrow lineshapes with tunable widths. Nanotechnology 17:2813–2820
12. Auguié B, Barnes WL (2008) Collective resonances in gold nanoparticle arrays. Phys Rev Lett 10:143902
13. Auguié B, Bendaña XM, Barnes WL, García de Abajo FJ (2010) Diffractive arrays of gold nanoparticles near an interface: Critical role of the substrate. Phys Rev B 82(15): 155447
14. Evlyukhin AB, Reinhardt C, Zywietz U, Chichkov B (2012) Collective resonances in metal nanoparticle arrays with dipole-quadrupole interactions. Phys Rev B 85(24):245411
15. Nikitin AG, Kabashin AV, Dallaporta H (2012) Plasmonic resonances in diffractive arrays of gold nanoantennas: near and far field effects. Opt Express 20(25):27941–27952
16. Sun C, Gao H, Shi R, Li C, Du C (2013) Design method for light absorption enhancement in ultra-thin film organic solar cells with the metallic nanoparticles. Plasmonics 8(2) 645–650
17. Hessel A, Oliner AA (1965). A new theory of Wood's anomalies on optical gratings. Appl Opt 4(10):1275–1297
18. Kravets VG, Schedin F, Grigorenko AN (2008) Extremely narrow plasmon resonances based on diffraction coupling of localized plasmons in arrays of metallic nanoparticles. Phys. Rev. Lett. 101(8):087403

19. Fedorenko L, Mamykin S, Lytvyn O, Burlachenko Y, Snopok B (2011) Nanostructuring of continuous gold film by laser radiation under surface plasmon polariton resonance conditions. Plasmonics 6(2):363–371.
20. Zhou W, Odom T (2011) Tunable subradiant lattice plasmons by out-of-plane dipolar interactions. Nat Nanotechnol 6:423–427
21. Vecchi G, Giannini V, Gomez Rivas J (2009) Surface modes in plasmonic crystals induced by diffractive coupling of nanoantennas. Phys Rev B 80:201401(R)
22. Offermans P, Schaafsma MC, Rodriguez SRK, Zhang Y, Crego-Calama M, Brongersma SH, Gómez Rivas J (2011) Universal scaling of the figure of merit of plasmonic sensors. ACS Nano 5(6):5151–5157
23. Rodriguez SRK, Lozano G, Verschuuren MA, Gomes R, Lambert K, De Geyter B, Hassinen A, Van Thourhout D, Hens Z Gómez Rivas J (2012) Quantum rod emission coupled to plasmonic lattice resonances: A collective directional source of polarized light. Appl Phys Lett 100(11):111103
24. Zhukovsky SV, Babicheva VE, Uskov AV, Protsenko IE, and Lavrinenko AV (2013) Electron Photoemission in Plasmonic Nanoparticle Arrays: Analysis of Collective Resonances and Embedding Effects. http://arxiv.org/abs/1308.3345
25. Ordal A, Bell RJ, Alexander Jr RA., Long LL, Querry MR (1985) Optical properties of fourteen metals in the infrared and far infrared: Al, Co, Cu, Au, Fe, Pb, Mo, Ni, Pd, Pt, Ag, Ti, V, and W. Appl Opt 24(24):4493–4499
26. CST Microwave Studio, http://www.cst.com/
27. Tamm I, Schubin S (1931) Zur Theorie des Photoeffektes an Metallen. Z Phys 68(1–2):97–113
28. Brodsky AM, Gurevich YY (1973) Theory of electron emission from metals, Nauka, Moscow
29. Brodsky AM, Gurevich YY (1968) Theory of external photoeffect from the surface of a metal. Sov Phys JETP 27:114–121
30. Scales C, Berini P (2010) Thin-film Schottky barrier photodetector models. IEEE J Quant Electron 46(5):633–643
31. West PR, Ishii S, Naik GV, Emani NK, Shalaev VM, Boltasseva A (2010) Searching for better plasmonic materials. Laser Photon Rev 4(6):795–808